\documentstyle[aps,prd,floats,epsf]{revtex}

\def\go{\mathrel{\raise.3ex\hbox{$>$}\mkern-14mu\lower0.6ex\hbox{$\sim$}}}
\def\lo{\mathrel{\raise.3ex\hbox{$<$}\mkern-14mu\lower0.6ex\hbox{$\sim$}}}
\def\be{\begin{equation}}
\def\ee{\end{equation}}

\def\sc{{{\rm sc}}}
\def\bk{{\bf k}}

\def\bo{{\bf \Omega}}

\newcommand{\beqa}{\begin{eqnarray}}
\newcommand{\eeqa}{\end{eqnarray}}
\newcommand{\cv}{c_V}
\newcommand{\ca}{c_A} 
\newcommand{\bhat}{\hat{\bf B}}
\newcommand{\bvec}{{\bf B}}
\newcommand{\kt}{T}
\newcommand{\fnu}{f_{\nu}}
\newcommand{\fN}{f_N}
\newcommand{\gnu}{g_{\nu}}
\newcommand{\bhnu}{{\bf h}_{\nu}}
\newcommand{\calbV}{{\cal \bf V}} 
\newcommand{\nubar}{\overline{\nu}}

\begin{document}

\twocolumn[\hsize\textwidth\columnwidth\hsize\csname
@twocolumnfalse\endcsname

\tightenlines
\draft
\title{ Can Parity Violation in Neutrino Transport Lead to 
Pulsar Kicks?}

\author{Phil Arras$^{1,2}$ and Dong Lai$^{1,3}$}
\address{$^1$Center for Radiophysics and Space Research, 
Cornell University, Ithaca, NY 14853}
\address{$^2$ Department of Physics, Cornell University}
\address{$^3$ Department of Astronomy, Cornell University} 

\date{\today} 
\maketitle 
\begin{abstract}

In magnetized proto-neutron stars, neutrino cross sections
depend asymmetrically on the neutrino momenta due to
parity violation. However, these asymmetric opacities do not
induce any asymmetric flux in the bulk interior of the star
where neutrinos are nearly in thermal equilibrium.
Consequently, parity violation in neutrino absorption and scattering 
can only give rise to asymmetric neutrino flux above the
neutrino-matter decoupling layer. The kick velocity is substantially reduced
from previous estimates, requiring a dipole field $B \sim 10^{16}$~G to get
$v_{\rm kick}$ of order a few hundred km~s$^{-1}$.

\end{abstract}
\bigskip

\vskip2pc]


Recent analyses of pulsar proper motion\cite{Lyne94}
indicate that neutron stars receive large kick
velocities at birth ($v_{\rm kick}=200-500$ km s$^{-1}$ on average, with
possibly a significant population having velocity $\go 1000$~km~s$^{-1}$).
Evidence for high velocity neutron stars has also come from observations
of pulsar bow shocks\cite{Cordes93} and studies of
pulsar-supernova remnant associations\cite{Frail94}.
Support for natal kicks has come from the detection of geodetic
precession in the binary pulsar PSR 1913+16\cite{Cordes90},
and orbital precession in the PSR J0045-7319 binary
and its fast orbital decay\cite{Kaspi96}.
In addition, evolutionary studies of the
neutron star binary population imply the
existence of pulsar kicks\cite{Fryer98}, and observations of nearby
supernovae and supernova remnants also support the notion that supernova
explosions are not spherically symmetric.

Mechanisms for the pulsar kicks may be divided into two classes.
The first class relies on hydrodynamical 
instabilities in the proto-neutron star\cite{Burrows95}
and nonspherical perturbations in the precollapse core\cite{Goldreich96}.
The asymmetries in the density and temperature
distributions naturally lead to asymmetric matter ejection and/or asymmetric
neutrino emission. 
In this paper, we are concerned with the second class of models,
where the kicks arise from asymmetric neutrino emission 
induced by strong magnetic fields.
The fractional asymmetry $\alpha$ in the radiated neutrino energy
required is $\alpha=Mv_{\rm kick}c/E_{\rm tot}$ ($=0.028$ for
$v_{\rm kick}=1000$~km~s$^{-1}$, neutron star mass
$M=1.4\,M_\odot$ and total neutrino energy radiated $E_{\rm tot}
=3\times 10^{53}$~erg).

A number of authors have noted that parity violation in weak
interactions can lead to asymmetric neutrino emission from
proto-neutron stars\cite{Chugai84,Dorofeev85,Vilenkin95}.
It has recently been suggested\cite{Horowitz97b}
that the asymmetry in neutrino emission may be enhanced due to multiple
scatterings of neutrinos by nucleons which are slightly polarized by the
magnetic field. Initial neutrino cooling calculations
\cite{Lai98a,Janka98} in magnetic fields appeared to indicate that a 
dipole field of order $10^{14}$~G is needed to produce kick velocity 
of $200$~km~s$^{-1}$. These results\cite{Horowitz97b,Lai98a,Janka98}
are wrong. Indeed, as we show in this paper,
although the neutrino scattering cross-section 
is asymmetric, detailed balance requires that 
there be no cumulative effect associated with multiple
scatterings in the bulk interior of the star where thermal equilibrium 
is maintained to a good approximation. 
Moreover, we derive an explicit expression for the neutrino flux 
in a magnetic field. In addition to the usual diffusive flux, there is 
a drift flux (along the magnetic field) which is 
proportional to the {\it deviation} of neutrino distribution from 
equilibrium. Hence parity violation can only induce an asymmetric neutrino flux
near the surface of the star. 

Our starting point is the Boltzmann transport equation for neutrinos 
(of a given species):
\beqa
\frac{ \partial \fnu(\bk)}{\partial t} + \bo \cdot \nabla \fnu(\bk)
& = & \left[{\partial f_\nu(\bk)\over\partial t}\right]_\sc
+ \left[{\partial f_\nu(\bk)\over\partial t}\right]_{\rm abs}
\label{boltzmanneqn}
\eeqa
where $\bk=k\bo$ is the neutrino momentum ($\bo$ is 
a unit vector), $\fnu(\bk)$ is
the neutrino distribution function (the position and 
time dependence are suppressed), and the scattering and absorption/emission 
collision terms are included on the right-hand side of the equation
(we set $\hbar=c=k_B=1$ throughout the paper).
In this paper we concentrate on neutrino scattering by nucleons ($\nu + N
\rightarrow \nu + N$) and electron neutrino absorption/emission from the
processes $n+\nu_{e} \rightleftharpoons p + e^{-}$, 
$p+\nubar_{e} \rightleftharpoons n + e^{+}$. Other processes
can be similarly considered, but are less important.

First we examine the scattering rate:
\beqa
&& \left[{\partial f_\nu(\bk)\over\partial t}\right]_\sc =
\sum_{ss'}\!\int\!\!\frac{d^3k'}{(2\pi)^3}\,\frac{d^3p}{(2\pi)^3}
\,\frac{d^3p'}{(2\pi)^3}\, W_{i\rightarrow f}^{(sc)}
\nonumber \\ && \times
\Bigl[(1-f_{\nu})(1-f_N)f_N'f_{\nu}'
- f_{\nu}f_N(1-f_N')(1-f_{\nu}') \Bigr],
\label{boltzmannscattering}
\eeqa
where $W_{i\rightarrow f}^{(sc)}=\left(2 \pi \right)^4 
\delta^4(P\!+\!K\!-\!P'\!-\!K'\!)
\left| M_{ss'}(\bo,\bo') \right|^2$ is S-matrix squared divided by time,
$P$ and $K$ ($P'$ and $K'$) are the 4-momenta of the initial (final) nucleon and
neutrino, $\fnu'\equiv \fnu(\bk')$ is the final state neutrino distribution 
function and  $f_N$ and $f_N'$ are the initial and final state
nucleon distribution functions, respectively.
Note that the energy of (nonrelativistic) 
nucleons includes the spin energy, $-s\mu_mB$
(where $s=\pm 1$ is the spin, $\mu_m$ is the magnetic moment)
\cite{protonnote}. 
In writing down eq.~(\ref{boltzmannscattering}) we have implicitly used
the relation $W_{i\rightarrow f}^{\rm (sc)}=W_{f\rightarrow i}^{\rm (sc)}$, 
reflecting the time reversal symmetry of the weak interaction. 

In the bulk interior of the proto-neutron star, neutrinos are 
to a good approximation in thermal equilibrium, with
$f_\nu\simeq f_\nu^{(0)}$ (the Fermi-Dirac function). 
Using the explicit forms for the neutrino and nucleon distribution
functions, together with energy conservation, we can verify the equality
\be
(1\!-\!f_{\nu}^{(0)})(1\!-\!f_N)f_N'f_{\nu}^{\prime (0)}
=f_{\nu}^{(0)}f_N(1\!-\!f_N')(1\!-\!f_{\nu}^{\prime (0)}),
\ee
representing detailed balance in thermal equilibrium.
Therefore the only nonzero contribution to 
$(\partial f_\nu/\partial t)_{\rm sc}$ must be proportional to the
deviation from thermal equilibrium; there can be no drift
flux along the magnetic field proportional to $f_\nu^{(0)}$. 

We now proceed to derive an expression for the neutrino flux. 
Using the Fermi-Dirac distribution functions of nucleons, 
we can write the scattering rate (\ref{boltzmannscattering}) in the form
\beqa
&& \left[{\partial \fnu(\bk)\over\partial t}\right]_\sc =
\int_0^{\infty}\!\! dk'\int\!\! d\Omega' \frac{ d\Gamma}{dk'd\Omega'}
\nonumber \\ && \times
\left[ e^{-q_0/\kt}(1-\fnu)\fnu'- \fnu (1-\fnu')
 \right]
\label{scatteringcollisionterm}
\eeqa
where $q_0=k-k'$ is the energy transfer, $d\Gamma/dk' d\Omega'$ is the 
differential cross-section (per unit volume) for scattering, as given by
\beqa
&& \frac{ d\Gamma}{dk'd\Omega'}  = 
\frac{ {k'}^2}{(2\pi)^3} \sum_{ss'}
\left| M_{ss'}(\bo,\bo') \right|^2 S_{ss'}(q_0,q)
\label{generalscatteringrate}
\eeqa
and the ``nucleon response function" is defined to be
\beqa
&& S_{ss'}(q_0,q) = 
\int\!\! \frac{ d^3p}{(2\pi)^3}\frac{ d^3p'}{(2\pi)^3}
\nonumber \\ && \times
\left(2 \pi \right)^4 \delta^4 \left(P+K-P'-K'\right)
\fN(1-\fN'),
\label{nucleonresponseinb}
\eeqa
where $q\equiv |\bk-\bk'|$.
Note that it is essential to retain the inelasticity in 
eq.~(\ref{scatteringcollisionterm}) in order to derive the neutrino 
drift flux. To lowest order in $1/m_N$, the matrix element for $\nu-N$ 
scattering\cite{antinote},
$\left| M_{ss'}(\bo,\bo') \right|^2$ is given by\cite{arraslai98}:
\beqa
&& \left| M_{ss'}(\bo,\bo') \right|^2  = 
\frac{1}{2} G_F^2 c_V^2 \left\{
\left( 1+3 \lambda^2 \right) + \left( 1-\lambda^2 \right) \bo \cdot \bo'
\right. \nonumber \\ && \left.
+ 2\lambda(\lambda+1)(s\bo+s'\bo')\cdot \bhat
- 2\lambda(\lambda-1)(s\bo'+s'\bo)\cdot \bhat
\right. \nonumber \\ && \left.
+ ss' \left[ \left( 1-\lambda^2 \right)(1+\bo \cdot \bo')
+4\lambda^2\bo \cdot \bhat \bo' \cdot \bhat \right]\right\}
\label{matrixelement}
\eeqa
where we have used the interaction Hamiltonian as given in 
Ref.~\cite{Raffelt96}, $G_F$ is the Fermi constant, 
$\ca$ and $\cv$ are the effective nucleon coupling constants, and 
$\lambda=\ca/\cv$.
Note that we have not summed over the initial or final state spins 
since both the nucleon distribution functions and the energy
conservation delta function depend on spins. Time-reversal symmetry,
$|M_{ss'}(\bo,\bo')|^2=|M_{s's}(\bo',\bo)|^2$, 
can be explicitly verified for the matrix element.
The nucleon response function for zero magnetic field
has been studied in Ref.~\cite{Reddy97} and 
can be directly generalized to the $B\neq 0$ case.  
Since the nucleon spin energy $\mu_m B$ is much smaller than other 
characteristic energies we may expand
eq.~(\ref{nucleonresponseinb}) to linear order in $B$.
Combining the expressions for $|M_{ss'}|^2$ and $S_{ss'}$ with
eq.~(\ref{generalscatteringrate}), we find\cite{arraslai98}
\beqa
&& \frac{ d\Gamma}{dk'd\Omega'}  =  A_0(k,k',\mu')
\nonumber \\ && 
+ \delta A_{+}(k,k',\mu') \bo \cdot \bhat
+ \delta A_{-}(k,k',\mu') \bo' \cdot \bhat
\label{expandedscatteringrate}
\eeqa
where $\mu'=\bo \cdot \bo'$. In (\ref{expandedscatteringrate}),
$A_0$ corresponds to the $\bvec=0$ result:
\beqa
A_0 \left(k,k',\mu' \right) &=&
\Lambda\left[\left(1+3\lambda^2\right)
+ \left( 1 -\lambda^2 \right)\mu'\right]\nonumber\\ 
&\times&\frac{1}{1-e^{-z}}
\ln\left( \frac{1+e^{-x_0}}{1+e^{-x_0-z}}\right)
\label{A0}
\eeqa
and the corrections arising from nonzero $\bvec$
involve the coefficients
\beqa
&&\delta A_{\pm} \left(k,k',\mu' \right) = 
\Lambda\left({2\lambda\mu_m B\over T}\right)
\frac{ 1}{\left(e^{x_0}+1 \right)}\nonumber \\ 
&&~~~~~~~~~\times\frac{ 1}{ \left( 1 + e^{-x_0-z} \right) }
\left( 1 \pm \lambda \frac{2m_Nq_0}{q^2} \right).
\label{Apm}
\eeqa
where we have defined 
\beqa
z&=&{q_0\over T},~~~\Lambda=\frac{{k'}^2}{(2\pi)^3}
\frac{G_F^2\cv^2m_N^2T}{\pi q},\\
x_0&=&{(q_0-q^2/2m_N)^2\over 4T(q^2/2m_N)}-{\mu_N\over T}.
\eeqa
The reason for writing the cross section in the form of
eq.~($\ref{expandedscatteringrate}$) is that the angular dependence
needed to find the moment equations is now manifest.
This differential cross section shows the signs of parity violation through
the $\bo\cdot \bhat$ and $\bo'\cdot \bhat$ terms. 
Typically, $\delta A_\pm$ is smaller than $A_0$ by 
a factor of order $\mu_mB/T$.

We now examine the macroscopic consequence of the asymmetric
cross section in eq.(\ref{expandedscatteringrate}). 
Let\cite{expand} 
\be
\fnu(\bk)=\fnu^{(0)}(k) + \gnu(k) + 3 \bo \cdot \bhnu(k),
\label{fexpand}\ee
where $\gnu$ is the deviation of the spherically symmetric part of $\fnu$ from 
the Fermi-Dirac value, while $\bhnu$ is the dipole dependent part which
leads to the flux\cite{noteflux}.
The first order moment equation is obtained by 
integrating eq.~(\ref{boltzmanneqn}) against $\int\!d\Omega \bo /4\pi$,
with the result
\beqa
&& \frac{ \partial \bhnu(k) }{\partial t}
+ \frac{1}{3}\nabla\left[ \fnu^{(0)}(k) + \gnu(k) \right]
\nonumber \\ &&
 =  \int \frac{d\bo}{4\pi} \bo 
\left\{ \left[{\partial f_\nu(\bk)\over\partial t}\right]_\sc
+ \left[{\partial f_\nu(\bk)\over\partial t}\right]_{\rm abs} \right\}.
\label{moment}
\eeqa
The scattering term yields
\beqa
&& \int \frac{d\Omega}{4\pi} \bo 
\left[{\partial \fnu(\bk)\over\partial t}\right]_\sc
= \calbV_0(k) + \delta \calbV_d(k).
\label{scmoment}\eeqa
Here
\beqa
\calbV_0(k) & = & 2\pi \int_0^{\infty}\!dk'\! \int_{-1}^{1} d\mu'
A_0(k,k',\mu')
\nonumber \\ && \times
\left[ \mu' \bhnu(k')C(k,k') + \bhnu(k)D(k,k') \right] 
\label{eqv0}
\eeqa
leads to the usual $B=0$ flux and the drift flux arises from the term
\beqa
&& \delta \calbV_d(k) = {2\pi\over 3} \bhat
\int_0^{\infty}\!dk'\!\int_{-1}^{1}\!d\mu'
\left[ \delta A_{+}(k,k',\mu')
\right. \nonumber \\ && \left.
+\mu' \delta A_{-}(k,k',\mu') \right]
\left[ \gnu(k')C(k,k') + \gnu(k)D(k,k') \right],
\label{eqvd}\eeqa
where
\beqa
C(k,k')  &=& e^{-q_0/T} \left(1-\fnu^{(0)} \right) + \fnu^{(0)},\\ 
D(k,k') &=& - \left[e^{-q_0/T}{\fnu^{(0)}}' +  1-{\fnu^{(0)}}' \right].
\eeqa

Since $g_\nu(k)$ is negligible in the bulk interior of the star, 
$\delta {\bf V}_d$ (and hence the drift flux) 
is only important outside the
neutrino decoupling sphere (outside of which there is no energy
exchange between neutrons and matter).
In this outer layer, the nucleons are nondegenerate and 
eqs.~(\ref{eqv0}) and (\ref{eqvd}) can be evaluated explicitly
under the condition $q_0/T\ll 1$. We find 
\beqa
&&\calbV_0(k)=-\kappa_0^{\rm (sc)}\bhnu(k),\label{eqv02} \\
&&\delta\!\calbV_d(k)\! =\! -{\kappa_0^{\rm (sc)}\!
\epsilon_{\rm sc}\over 3}\!
\left[\gnu\!(k)\! + \!
\kt\! \frac{\partial\gnu\!(k)}{\partial k}\! 
-\! 2\gnu\!(k)\!\fnu^{(0)}\!(k)\!\right]\!
\bhat\!
\label{eqvd2}
\eeqa
where $\kappa_0^{\rm (sc)}$ is the zero-field scattering opacity (per unit 
volume), as given by
\be
\kappa_0^{\rm (sc)}={8\pi\over 3} \left( \frac{G_F \cv k}{2\pi} \right)^2
\left( 1 + 5\lambda^2 \right) n_N,
\ee
($n_N$ is nucleon number density), and the asymmetry parameter is given by
\be
\epsilon_{\rm sc}=
{6\lambda^2\over (1+5\lambda^2)}\frac{\mu_mB}{\kt}\simeq {\mu_mB\over T}.
\ee
Note that for nondegenerate nucleons $q_0/T \sim k/(m_NT)^{1/2} \ll 1$.
Naively one would expect from eq.~(\ref{scatteringcollisionterm})
or eq.~(\ref{eqvd}) that $\delta {\bf V}_d$ would be suppressed by
a factor of order $q_0/T$\cite{q0note}. 
However, inspection of eq.~(\ref{Apm}) shows that $A_{\pm}$ contains the term
$2m_Nq_0/q^2$ which can be quite large and has the same parity (around 
$k'=k$) as $q_0/\kt$. The result is that
$\delta \calbV_d$ in eq.~(\ref{eqvd2})
is not suppressed by a factor of $q_0/\kt$ outside
the neutrino decoupling sphere. 

The zeroth moment of eq.~(\ref{scatteringcollisionterm}) 
describes the energy exchange between matter and neutrino fields. 
The asymmetric scattering due to the magnetic
field modifies this equation, but this only affects the 
rate at which the neutrinos approach equilibrium.

We now consider the effect of asymmetric absorption/emission. 
For concreteness, we focus on the process $\nu_e+n\rightleftharpoons p+e^-$.
The absorption/emission rate can be
written as
\begin{eqnarray}
&&\left[{\partial f_\nu(\bk)\over\partial t}\right]_{\rm abs}
=\sum_{s_ns_p}\!\int\!\!{d^3p_n\over (2\pi)^3}
{d^3p_p\over (2\pi)^3}\sum_eW_{i\rightarrow f}^{\rm (abs)}\nonumber\\
&&\times\Bigl[f_ef_p(1-f_\nu)(1-f_n)
-f_\nu f_n(1-f_p)(1-f_e)\Bigr],
\label{absorb}\end{eqnarray}
where $s_n,\,s_p$ are the neutron and proton spins,
$\sum_e$ stands for summing over the electron
phase space\cite{Landaunote} and
$W^{\rm (abs)}\propto |M^{\rm (abs)}|^2\delta(E_n+E_\nu-E_p-E_e)$
is the transition rate for absorption (with matrix element $M^{\rm (abs)}$).
Expanding $f_\nu(\bk)$ as in eq.~(\ref{fexpand}), 
and using the equality
\be
{(1-f_n)f_pf_e\over f_n(1-f_p)(1-f_e)}=e^{(\mu_\nu-k)/T},
\ee
where $\mu_\nu\equiv\mu_p+\mu_e-\mu_n$, we find
\be
\left[{\partial f_\nu(\bk)\over\partial t}\right]_{\rm abs}
=-\left[g_\nu(k)+3\bo\cdot\bhnu(k)\right]
\kappa^{\rm (abs)},
\ee
where the absorption opacity is given by 
\begin{eqnarray}
\kappa^{\rm (abs)}&=&\left[1+e^{(\mu_\nu-k)/T}\right]
\sum_{s_ns_p}\!\int\!\!{d^3p_n\over (2\pi)^3}
{d^3p_p\over (2\pi)^3}\sum_eW_{i\rightarrow f}^{\rm (abs)}\nonumber\\
&\times& f_n(1-f_p)(1-f_e).
\end{eqnarray}
Calculation\cite{Dorofeev85,arraslai98} 
shows that $\kappa^{\rm (abs)}$
can be expressed in terms of the $B=0$ opacity
$\kappa_0^{\rm (abs)}$ via
\be
\kappa^{\rm (abs)}=\kappa_0^{\rm (abs)}(1+\epsilon_{\rm abs}{\bf \Omega}\cdot\bhat).
\ee
Taking the first moment of eq.~(\ref{absorb}), we find
\beqa
\int\!{d\Omega\over 4\pi}{\bf\Omega}
\left[{\partial f_\nu(\bk)\over\partial t}\right]_{\rm abs}
&& =-\kappa^{\rm (abs)}_0{\bf h}_\nu(k)
\nonumber \\ && 
-{1\over 3}\epsilon_{\rm abs}\kappa_0^{(\rm abs)}g_\nu(k)\bhat.
\label{absmoment}\eeqa
This clearly indicates that in the bulk interior of the star, where
$g_\nu\simeq 0$, there is no drift flux associated with asymmetric
absorption; the asymmetry from neutrino absorption 
is exactly cancelled by that from emission\cite{absnote}.
 
Combining eqs.~(\ref{moment}),~(\ref{scmoment}),~(\ref{eqv02}),~(\ref{eqvd2}),
and (\ref{absmoment}), we find the explicit expression for 
the neutrino flux in the outer layer of the star:
\beqa
&&{\bf h}_\nu(k)\!=\!-{1\over 3\kappa_0^{(\rm tot)}}
\nabla \left[f_\nu^{(0)}(k)+g_\nu(k)\right]\!
-\!{\epsilon_{\rm abs}\over 3}{\kappa_0^{\rm (abs)}\over
\kappa_0^{\rm (tot)}}g_\nu(k)\bhat \nonumber\\
&& ~~-{\epsilon_{\rm sc}\over 3}
{\kappa_0^{\rm (sc)}\over\kappa_0^{\rm (tot)}}
\left\{\gnu(k)\left[1-2\fnu^{(0)}(k)\right]+
\kt \frac{\partial \gnu(k)}{\partial k}\right\}\bhat,
\label{flux}\eeqa
where $\kappa_0^{\rm (tot)}(k)=\kappa_0^{\rm (sc)}+\kappa_0^{\rm (abs)}$,
and we have neglected $\partial\bhnu/\partial t$.
Equation (\ref{flux}) clearly shows that the neutrino drift flux
(the terms with $\bhat$) is proportional to $g_\nu$, which 
decreases rapidly with increasing optical depth.
This indicates that asymmetric neutrino 
opacities due to parity violation do not affect neutrino transport
in the bulk interior of the star. Asymmetry in
neutrino flux can only arise from regions above the neutrino 
decoupling layer. 

Clearly, to determine the asymmetry in neutrino emission from a magnetized
proto-neutron star requires knowledge of the temperature
profile and the function $g_\nu(k)$ near the stellar surface. In principle, 
one can use the moment equations to calculate $g_\nu(k)$, 
but the temperature profile must also be determined self-consistently.
A full neutrino transport calculation is beyond the scope of this paper.
We now give a rough estimate of the neutrino asymmetry
due to the $\nu_e,~\bar\nu_e$ drift flux
(the net drift flux associated with $\mu$ and $\tau$ neutrinos is
zero\cite{antinote}). The asymmetry parameter for scattering
is of order $\epsilon_{\rm sc}\sim 0.006B_{15}/T$, where 
$B_{15}$ is the field strength in units of $10^{15}$~G, and 
$T$ the temperature in MeV. 
The asymmetry parameter for absorption, $\epsilon_{\rm abs}$, 
is dominated for low energy neutrinos ($\lo 15$~MeV)
by the contribution from electrons in the ground Landau 
level\cite{Dorofeev85,arraslai98} 
\be
\epsilon_{\rm abs}\simeq 0.1{eB\over (E_\nu+Q)^2}
\simeq 0.6\,B_{15}E_\nu^{-2},\ee
(where $E_\nu$ is the neutrino energy in MeV, $Q=1.29$~MeV), 
and for high energy neutrinos by nucleon polarization ($\sim \mu_mB/T$).
The electron neutrinos decouple from matter near the neutrinosphere, 
where typical density and temperature are $\rho\sim 10^{12}$~g~cm$^{-3}$, 
and $T\sim 3$~MeV. For a mean $\nu_e$ energy of $10$~MeV, $\epsilon_{\rm abs}$ 
is greater than $\epsilon_{\rm sc}$. The asymmetry in the $\nu_e$,
$\bar\nu_e$ flux is approximately given by the ratio of
the drift flux and the diffusive flux, of order
$\epsilon_{\rm abs} [\kappa_0^{\rm (abs)}/\kappa_0^{\rm (tot)}]$.
Averaging over all neutrino species, we find the total asymmetry
in neutrino flux $\alpha\sim 0.2\epsilon_{\rm abs}$.
To generate a kick of a few hundreds per second would require a
dipole field of order $10^{16}$~G.

After our paper was largely completed, a preprint by 
Kusenko et al.\cite{Kusenko98} came to our attention, where the
authors came to a similar conclusion. However, they only showed
that there is no flux in equilibrium, and did not derive an expression
for the flux. In addition, they contend that unitarity necessarily
implies that the cross section cannot depend asymmetrically on the 
neutrino momenta, which is false (see eq. 
($\ref{expandedscatteringrate}$)).

We thank O. Grimsrud, Y.-Z. Qian, and I. Wasserman for useful discussion.
D.L. acknowledges support from the Alfred P. Sloan foundation.



\begin{references}

\bibitem{Lyne94}
A.G. Lyne and D.R. Lorimer, Nature, {\bf 369}, 127 (1994);
B.M.S. Hansen and E.S. Phinney, MNRAS, {\bf 291}, 569 (1997);
J.M. Cordes and D.F. Chernoff, ApJ, in press (1997). 

\bibitem{Cordes93}
J.M. Cordes, R.W. Romani and S.C. Lundgren, Nature, {\bf 362}, 133 (1993).

\bibitem{Frail94}
D.A. Frail, W.M. Goss and J.B.Z. Whiteoak, ApJ,
{\bf 437}, 781 (1994).

\bibitem{Cordes90}
J.M. Cordes, I. Wasserman, and M. Blaskiewicz, ApJ, {\bf 349}, 546 (1990);
M. Kramer, ApJ, submitted (1998). 
 
\bibitem{Kaspi96}
V.M. Kaspi, et al., Nature, {\bf 381}, 583 (1996);
D. Lai, ApJ, {\bf 466}, L35 (1996).
 
\bibitem{Fryer98}
e.g., C. Fryer, A. Burrows, and W. Benz, ApJ, in press (1998).

\bibitem{Burrows95}
e.g., A. Burrows, J. Hayes and B.A. Fryxell, ApJ, {\bf 450}, 830;
H.-T. Janka and E. M\"uller, A\&A, {\bf 306}, 167 (1996);
M. Herant, et al., ApJ, {\bf 435}, 339 (1994).
 
\bibitem{Goldreich96}
P. Goldreich, D. Lai and M. Sahrling, in
{\it Unsolved Problems in Astrophysics}, ed. J.~N. Bahcall and
J.~P. Ostriker (Princeton Univ. press);
A. Burrows and J. Hayes, Phys. Rev. Lett. {\bf 76}, 352 (1996).
 
\bibitem{Chugai84}
N.N. Chugai, Sov. Astron. Lett. {\bf 10}, 87 (1984).
 
\bibitem{Dorofeev85}
O.F. Dorofeev, et al., Sov. Astron. Lett. {\bf 11}, 123 (1985).
 
\bibitem{Vilenkin95}
A. Vilenkin, ApJ, {\bf 451}, 700 (1995).
 
\bibitem{Horowitz97b}
C.J. Horowitz and G. Li, Phys. Rev. Lett. {\bf 80}, 3694 (1998).
 
\bibitem{Lai98a}
D. Lai and Y.-Z. Qian, ApJ, {\bf 495}, L103 (1998);
erratum (July 1, 1998 ApJ Lett.).
 
\bibitem{Janka98}
H.-T. Janka, in {\it Proceedings ``Neutrino Astrophysics''}, ed.
M. Altmann et al. (Tech. Univ. M\"unchen, Garching, 1998) 
(astro-ph/9801320).

\bibitem{protonnote}
We treat the proton's motion as that of a free particle instead of 
using Landau levels. Since many levels are occupied for the conditions 
in a proto-neutron star, the effect of level quantization is negligible.
 
\bibitem{antinote}
Equation (\ref{matrixelement}) applies for neutrino. Because of
the crossing symmetry, one can obtain 
the expression for $\bar\nu$ by switching $\bo$ and $\bo'$.
Similarly, one switches $\bo$ and $\bo'$ in 
eq.~(\ref{expandedscatteringrate}) to obtain the cross section for
$\bar\nu$.

\bibitem{arraslai98}
P. Arras and D. Lai, to be submitted to Phys. Rev. D (1998).

\bibitem{Raffelt96}
G. Raffelt, {\it Stars As Laboratories for Fundamental Physics},
(The Univ. of Chicago Press: Chicago, 1996).

\bibitem{Reddy97}
S. Reddy, M. Prakash and J.M. Lattimer, astro-ph/9710115.
 
 
\bibitem{expand}
Note that in principle, there is a quadrupole term which 
contributes to the drift flux. This term is similar to $g_\nu(k)$,
although their dependence on optical depth may differ. For simplity we
have neglected the quadrupole term. 

\bibitem{noteflux} The specific neutrino energy flux ${\bf F}_\nu(k)$
is related to $\bhnu$ by ${\bf F}_\nu(k)=k^3\bhnu(k)/(2\pi)^3$.

\bibitem{q0note} In eq.~(\ref{eqvd}), 
since $C(k,k')\simeq 1+{\cal O}(q_0/T)$ and
$D(k,k')\simeq -1+{\cal O}(q_0/T)$, we finds 
that $[g_\nu(k')C+g_\nu(k)D]$ is proportional to $q_0/T$.
The same holds for the factor inside the square bracket of 
eq.~(\ref{scatteringcollisionterm}).


\bibitem{Landaunote}
This includes summing over the Landau levels, spin, and the $z$-momentum.
Note that transverse momentum is not conserved in magnetic
fields.
 
\bibitem{absnote}
This was already pointed out in Ref.~\cite{Lai98a},
although Kirchhoff's law was assumed in arriving at this result.
 
\bibitem{Kusenko98} 
A. Kusenko, G. Segre and A. Vilenkin, astro-ph/9806205.



\end{references}
\end{document}